\documentclass[twocolumn]{aastex62}

\usepackage{natbib}
\usepackage[style=american]{csquotes}
\usepackage{epsfig}
\usepackage{apjfonts}
\usepackage{lmodern} 
\usepackage{graphicx}
\usepackage{tabularx}   
\usepackage{xcolor}   
\usepackage{mathrsfs}
\usepackage{amsmath,amssymb}
\usepackage{verbatim}
\usepackage{wrapfig}
\usepackage{enumitem}
\usepackage{longtable}
\usepackage{bm}
\usepackage{threeparttable}
\usepackage{booktabs}
\usepackage{multirow}
\usepackage{comment}
\usepackage{hyperref}
\usepackage{multirow}

\definecolor{cerulean}{rgb}{0,0.482,0.655}

\definecolor{purple}{rgb}{0.627, 0.125,0.941}

\def\co2{{CO$_2$}}
\def\h2o{{H$_2$O}}
\def\ch4{{CH$_4$}}

\accepted{October 31, 2022}
\submitjournal{ApJ}

\shorttitle{Sample Selection in Exoplanet Atmosphere Population Studies}
\shortauthors{Batalha, Wolfgang, Teske et al.}
\graphicspath{{./}{figures/}}

\begin{document}

\title{Importance of Sample Selection in Exoplanet Atmosphere Population Studies}

\author[0000-0003-1240-6844]{Natasha E. Batalha}
\affiliation{NASA Ames Research Center, Moffett Field, CA 94035, USA}

\author[0000-0003-2862-6278]{Angie Wolfgang}
\affiliation{Senior Data Scientist, SiteZeus}

\author{Johanna Teske} 
\affiliation{Earth and Planets Laboratory, Carnegie Institution for Science, 5241 Broad Branch Road, NW, Washington, DC 20015, USA}

\author[0000-0003-4157-832X]{Munazza K. Alam}
\affiliation{Earth and Planets Laboratory, Carnegie Institution for Science, 5241 Broad Branch Road, NW, Washington, DC 20015, USA}

\author[0000-0001-8703-7751]{Lili Alderson} 
\affiliation{School of Physics, University of Bristol, HH Wills Physics Laboratory, Tyndall Avenue, Bristol BS8 1TL, UK}

\author[0000-0002-7030-9519]{Natalie M. Batalha}
\affiliation{Department of Astronomy and Astrophysics, University of California, Santa Cruz, CA 95064, USA}

\author[0000-0003-3204-8183]{Mercedes L\'opez-Morales} 
\affiliation{Center for Astrophysics ${\rm \mid}$ Harvard {\rm \&} Smithsonian, 60 Garden St, Cambridge, MA 02138, USA}

\author[0000-0003-4328-3867]{Hannah R. Wakeford} 
\affiliation{School of Physics, University of Bristol, HH Wills Physics Laboratory, Tyndall Avenue, Bristol BS8 1TL, UK}



\begin{abstract}
Understanding planet formation requires robust population studies, which are designed to reveal trends in planet properties. In this work, we aim to determine if different methods for selecting populations of exoplanets for atmospheric characterization with JWST could
influence population-level inferences. We generate three hypothetical surveys of super-Earths/sub-Neptunes, each spanning a similar radius-insolation flux space. The survey samples are constructed based on three different selection criteria (evenly-spaced-by-eye, binned, and a quantitative selection function). Using an injection-recovery technique, we test how robustly individual-planet atmospheric
parameters and population-level parameters can be retrieved. We find that all three survey designs result in 
equally suitable targets for individual atmospheric characterization, but not equally suitable targets for constraining population parameters.
Only samples constructed with a quantitative method or that are sufficiently evenly-spaced-by-eye result in robust population parameter constraints. Furthermore, we find that the sample with the best targets for individual atmospheric study does not necessarily result in the best constrained population parameters. The method of sample selection must be considered.
We also find that there may be large variability in population-level results with a sample that is small enough to fit in a single JWST cycle ($\sim$12 planets), suggesting that the most successful population-level analyses will be multi-cycle. Lastly, we infer that our exploration of sample selection is limited by the
small number of transiting planets with measured masses around bright stars. Our results can guide future development of programs that
aim to determine underlying trends in exoplanet atmospheric properties and, by extension, formation and evolution processes.

\end{abstract}

\keywords{Exoplanet atmospheres -- Hierarchical models -- Surveys}

\section{Introduction \& Motivation} \label{sec:intro}
One of the primary drivers in studying exoplanets is to leverage their atmospheres to understand the origin and formation of planetary systems. Developing a comprehensive understanding of planet formation requires going beyond characterizing individual systems to conducting robust population studies of larger samples where trends can be revealed \citep{Bean2017statistical}. In these studies it is important to consider the multiple levels of decisions and biases built into single or ensemble observations. 
Biases that arise from decisions regarding which targets are detected, which targets are followed-up, and how they are ultimately observed are often not considered. For example, the diverse interests and priorities of observers may have influenced the inferred boundaries of the ``brown dwarf desert'' such that companions slightly above the planetary-mass regime were not followed up or published until recently \citep[e.g.][]{Kiefer2019}. In principle these types of biases can be accounted for, but this requires careful documentation about both telescope/instrument performance and how observations were conducted. Unfortunately the latter is often not available, and thus inferences from population studies can be flawed (e.g., \citealt{montet2018, Burt2018}). 
Importantly, failure to account for how certain targets were chosen over others prevents accurate inference of the population distribution from the observed sample. For example, newly detected transiting planets are often chosen for mass measurement because they are novel in some way.  This decision biases our understanding of the full distribution of planet bulk densities toward the extremes; the very planets that should drive population trends via their ``ordinariness'' are systematically missing or poorly constrained. As another example, a ranking based on scale height (e.g. TSM in \citet{kempton2018tsm}) might bias a sample towards high equilibrium temperature, and/or low gravity planets.

JWST will, for the first time, make possible population studies of the atmospheres of planets smaller than Neptune, where there are currently only a handful of atmospheric measurements \citep{kreidberg2014,Tsiaras201655cnce,Guo2020,Benneke2019k218,mikal2020,Guilluy2020,Libby2021}. While the Solar System reflects a bimodality in planet and atmosphere types -- gas giant/primary atmospheres and terrestrial/secondary atmospheres -- outside the Solar System this bimodality seems to be blurred 
especially for planets intermediate in size between terrestrial and gas giant, which are also the most common type of planet at periods $\lesssim 100$ days \citep{Fulton2017,Fulton2018CKS}. Thus it is of great interest whether close-in ``sub-Neptune'' and ``super-Earth'' planets have primordial atmospheres dominated by H/He and/or secondary atmospheres outgassed from their interiors, and whether there exists a transition between them. Though this question is complex and multi-faceted, an initial investigation could start with determining  whether or not there is an observable transition in atmospheric composition that accompanies varying planet radii and stellar irradiation (referred to hereafter as flux).

In this work we aim to determine if and/or 
how different methods for selecting populations of exoplanets for atmospheric characterization could influence population-level inferences about a composition-radius-flux relation.  We first generate three samples of planet based on different selection criteria (\S\ref{sec:targets}) 
to create hypothetical JWST NIRSpec surveys, each consisting of 12 planets. We chose NIRSpec because it combines excellent detector performance \citep{Birkmann2022Nirspec} with the high resolution and large wavelength coverage modes optimal for studies of exoplanet atmospheres \citep{BatalhaLine2017}.

Next, we define a fiducial population relation and apply it to all three samples, which allows us to assign an atmospheric composition to each individual planet and thus simulate an ``observed'' spectrum for each  (\S\ref{sec:simulation}). 
We then ``retrieve'' the atmospheric properties of the individual planets to back out the injected population relation from each simulated survey via a hierarchical Bayesian model (\S\ref{sec:hbm}). Lastly, we present a comparison of the inferred individual-planet atmospheric parameters and population-level parameters between the three samples (\S\ref{sec:results}). Ultimately, our results can help guide future planning for exoplanet atmosphere observations, which we summarize in (\S\ref{sec:summary}).

\section{Target Sample Selection} \label{sec:targets}

We choose to simulate three samples based on different selection criteria. The first ``quantitatively selected'' (QS) sample represents an approach to sample selection that is designed for population analysis, wherein a quantitative merit function is used to choose targets. Specifically, we use the methodology that was the basis of the selected large JWST program PID\# 2512. The targets are a subset of the $R_p \leq 3$~R$_{\oplus}$ planets observed as part of the Magellan-TESS Survey (MTS), which itself was created using a quantitative ranking function based on $R_p$, insolation flux, and the expected observing time required to reach 2 m~s$^{-1}$ photon-limited RV precision (which is a function of $V$ mag and spectral type) applied to TOIs detected during Year 1 of TESS (see \citealt{Teske2021} for details). All of the planets in the final MTS sample are guaranteed to have mass constraints from a homogeneous analysis, which is why we opted to keep the QS sample a subset of the MTS targets. 
The second ``three bin'' (3Bin) and third ``evenly space by eye'' (ESBE) samples, on the other hand, are both drawn from confirmed planets with $R_p \leq 3$~R$_{\oplus}$ and measured masses, retrieved from the NASA Exoplanet Archive \citep{ps}\footnote{in March 2021, \texttt{https://exoplanetarchive.ipac.caltech.edu}}. We do not set a mass precision requirement, only that the mass is not an upper limit. 

We acknowledge that by limiting the QS sample to a subset of the MTS targets, we potentially hinder this sample's statistical power. 
However, for this analysis we opt to test the choices made for the upcoming JWST program specifically, where it was more important to have verified targets with mass constraints from a homogeneous analysis. Additionally, in \S\ref{sec:results} we demonstrate that all three samples have an 
opportunity to ``succeed'' based on how well we are able to retrieve atmospheric and/or population parameters.

\subsection{Quantitatively Selected Sample (QS)} \label{subsec:MTS}

\begin{figure*}[htbp]
\centering
\includegraphics[width=\textwidth]{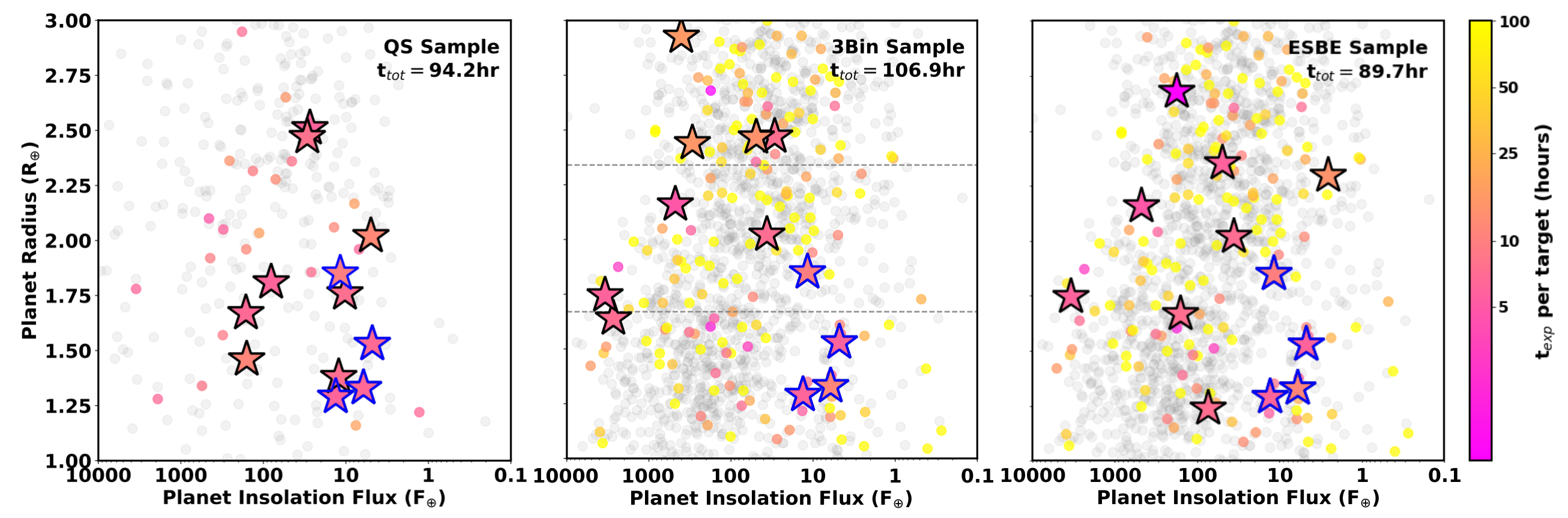} 
\caption{The radii and insolation fluxes for the three planet samples with which we conducted our survey simulation. The large bold stars show the targets selected in each sample; those outlined in blue happen to be included in all three samples. The small colored circles show the starting population from which these targets were selected before any filtering (\S2). The color of the symbols corresponds to the estimated JWST $t_{\rm exp}$ as described in the text; grey symbols represent all year 1 TOIs (QS plot) or planets without mass measurements in the NASA Exoplanet Archive (3Bin and ESBE plots). The dashed lines in the middle (3Bin) panel indicate the bin spacing as described in the text. } 
\label{fig:samples}
\end{figure*}

To select a subset of the MTS targets, we explore different ranking metrics using planet insolation flux $F_{\rm{insol}}$, planet radius $R_p$, host star effective temperature $T_{\rm{eff}}$, and the JWST integration time $t_{\rm exp}$, with the goal of addressing the population-level question: \textit{Is there an observable transition in atmospheric composition that accompanies varying planet radii and flux?} We choose a ranking metric that results in a sample with less variation in $T_{\rm{eff}}$ (a rough proxy for the high-energy radiation environment of the planet: \citealt{Linsky2014}) and $F_{\rm{insol}}$ (see Figure \ref{fig:samples}). Our final, purely empirical merit function applied to down-select the MTS sample is:




\begin{equation} \label{eqn:merit2}
\begin{split}
\textrm{merit}_{\rm{mock-atm}} = & t_{\rm{exp}}^{-2} \times \\
&\mathcal{N}(R_p;1.7,1) \times \\
&\mathcal{N}(T_{\rm{eff}};4000,200) \times \\
&\mathcal{N}(\rm{log}(F_{\rm{insol}});1.5,0.3)
\end{split}
\end{equation}

\noindent where $\mathcal{N}$(variable; $\mu,\sigma$) represents a normal distribution with mean $\mu$, standard deviation $\sigma$, and lower limit of 0, in the same units as the corresponding variable (R$_\oplus$, K, and log(F$_\oplus$), respectively). For each target, the JWST $t_{\rm exp}$ is calculated as the time required to achieve a 30~ppm spectral precision sampled at $R=100$ with NIRSpec G395H at 4~$\mu$m. We use \texttt{PandExo} \citep{batalha2017pandexo} to compute an optimal duty cycle for each observation, where each transit event is assumed to have a total time of twice the transit duration. The strength of the dependency on $t_{\rm{exp}}$ of $-2$  was chosen on a trial-and-error basis to ensure a balanced JWST large program of a reasonable size (the two approved JWST large GO programs for transiting exoplanet science are 142 and 75 hours). If the $t_{\rm exp}$ exponent was too steep, the sample would be biased towards bright targets, with short transit durations (which translates to short orbital periods). If the exponent was too shallow, the program size would not be feasible within a single large program. A value of $-2$ offered a balance between these two end-cases.  Though it is beyond the scope of the analysis, exploring how this exponent affects the chosen sample, and the results of the population analysis, is an important question that we leave to future work.

Applying this ranking metric to the MTS list as of April 2020 results in the following TOIs (in rank order): 260.01, 776.02, 836.01, 562.01, 134.01, 175.01, 836.02, 687.01, 776.01, 455.01, 175.02, 186.01, 174.01, 174.02, 402.01, 402.02. We removed TOI 687.01 due to the uncertain period of the planet, and TOI 186.01 because it would saturate the NIRSpec detector. We remove TOI 174.01 and .02 because they have not been validated as a confirmed planets. 
This results in a sample of 12 targets as 
shown in Figure \ref{fig:samples} (left panel). We fix the number of targets to twelve such that it could feasibly fit into a single large JWST program. Though quantifying 
how well population parameters can be retrieved as a function of sample size would be an important exploration, 
it is beyond the scope of this analysis. The total exposure time for this sample ($\sum_{n=1}^{12}t_{\rm exp} $) is 94.2 hours.

\subsection{Three Bin Sample (3Bin)} \label{subsec:3Bin}

Our second sample represents a binned approach to sample selection, wherein planets of different sizes are equally represented on the target list and are otherwise prioritized by how easy they are to observe. 
For each planet we calculate a JWST integration time $t_{\rm exp}$ using the same formulation as the QS sample. Then we separate these planets into three $R_p$ bins: $1.0-1.67$, $1.67-2.34$, and $2.34-3.0$ R$_{\oplus}$. Within each bin we rank the targets by their $t_{\rm exp}$ (discounting targets that would saturate the detector by exceeding 100\% full well within two frame times, known as a ``hard saturation''), and fill in a list of 12 targets by selecting the top four targets in each bin. The radius bins are roughly chosen to represent planets that are rocky, planets that are likely sub-Neptunes, and those that are in between. 

We also tried implementing an insolation flux dimension (4 flux~$\times$~3 radius bins), such that there was one target per bin. We first attempted a log-normal spacing between 4-4000~$F_\odot$. However, the ``best'' target  in the highest insolation flux \& radius bin (K2-66 b) has $t_\mathrm{exp}=$101~hr resulting in an unfeasible total sample time of 201~hrs. We also tried extending the bounds of the insolation flux parameter space to cover both lower and higher fluxes. Each of these attempts resulted in samples that had at least two empty radius-flux bins (out of 12 total). In those cases, additional questions arose regarding how to reassign targets. For example, if one radius-flux bin is empty is it more optimal to assign a target to an adjoining bin? Or, is it better to choose the next highest ranked target (e.g., by $t_\mathrm{exp}$) regardless of what bin it may fall in? Though these are interesting questions, addressing them  was ultimately beyond the scope of the analysis in this paper.

Therefore, we ultimately opted for a binning method with a single dimension (radius). We emphasize that this method has the potential to create a highly skewed sample (e.g., if all the targets in a single radius bin fell into a narrow flux range). Our 3Bin sample selection method naturally resulted in a sample that was fairly evenly-spaced in flux. Overall, we note that this ``missing-bin'' problem is common when the number of targets with mass constraints (colored points in Figure \ref{fig:samples}, middle and right) is small, and therefore we view our choice as a viable test for this analysis. 

We choose the same number of planets as the QS sample to ensure that our population relation retrieval results is not driven by differences in sample size. 
The resulting 3Bin targets, shown in Figure \ref{fig:samples} (middle panel) include GJ 357 b (TOI-562.01), LTT 1445 A b (TOI-455.01), L 98-59 d (TOI-175.02), Kepler-21 b, HD 86226 c (TOI-652.01), HD 213885 b (TOI-141.01), GJ 9827 d, TOI-776 b, HD 15337 c (TOI-402.02), HIP 116454 b, HD 106315 b, and TOI-824 b. We note that five of the targets in this sample overlap with the QS 
sample. 
The total exposure time for the 3Bin sample is 106.9 hours.

\subsection{Evenly-Spaced-by-Eye Sample (ESBE)} \label{subsec:ESBE}

Our third sample represents an approach to sample selection wherein targets are hand-picked to evenly cover the $R_{pl}-F_{insol}$ plane. Hand-picking has the potential to prioritize planets that are novel or unusual in terms of their size or insolation flux.  
Since we have no direct ranking on observing feasibility, we filter the sample to include only targets with $t_{\rm exp} \leq 15$ hours (discounting targets that would hard saturate within two groups), and plot the planet $R_p$ and $F_{\rm{insol}}$ values. A fifteen hour $t_{\rm exp}$ limit ensures no one planet dominates the total time allocated to the program. Then we choose a sample of 12 targets (again, the same size as our QS sample) across this parameter space. The ESBE targets, shown in Figure \ref{fig:samples} (right panel), are TOI-421 b, HD 97658 b, LTT 3780 c (TOI-732.02), HD 213885 b (TOI-141.01), HD 86226 c (TOI-652.01), GJ 9827 d, TOI-776 b, HD 15337 b (TOI-402.01), GJ 9827 c, GJ 357 b (TOI-562.01), LTT 1455 A b (TOI-455.01), and L 98-59 d (TOI-175.02). We note that five (seven) of the targets in this sample overlap with the QS (3Bin) samples, respectively, and that four planets are common across all three samples (see Figure \ref{fig:samples}). The total exposure time for the ESBE sample is 89.7 hours.

\section{Survey Simulation} \label{sec:simulation}

\begin{figure*}[htbp]
\centering
\includegraphics[width=\textwidth]{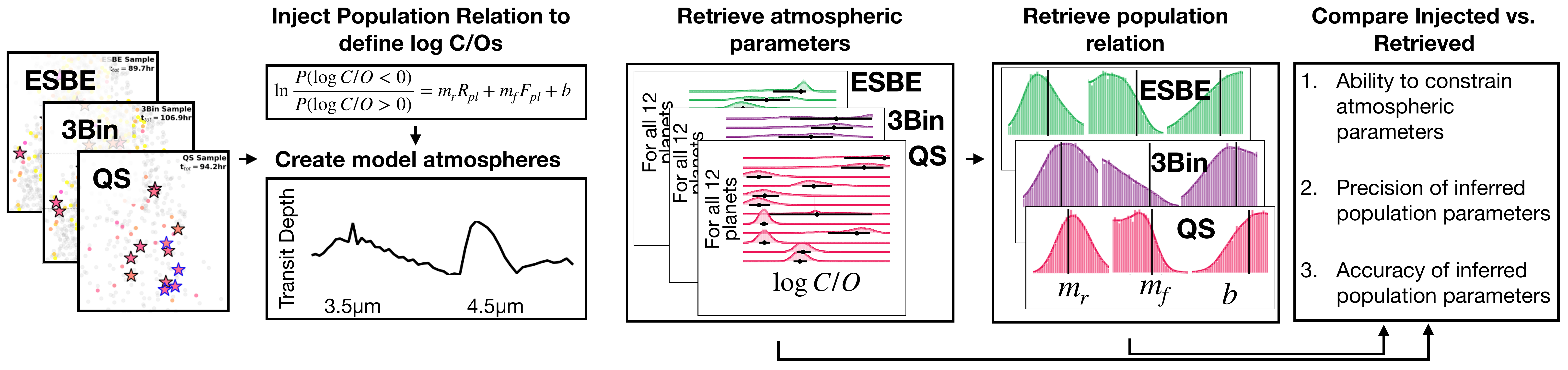} 
\caption{This flowchart outlines the steps in our simulation study, which assesses how sample selection affects inferences about populations of exoplanet atmospheres. We first generate three 12-planet samples using different approaches to target selection. We inject the same fiducial population relation into each sample to then create simulated spectral observations. Next we retrieve atmospheric parameters for each planet and then use them to infer the population relation that would be derived from each of the three samples. In the end we determine how each sample performed by computing the accuracy and precision of both the retrieved atmosphere and the population parameters.} 
\label{fig:flowchart}
\end{figure*}

With our three 12-planet samples in hand, we can next proceed with simulating populations of planetary atmospheres.  In particular, we are interested in how sample selection affects the ability to determine population-level inferences regarding composition-radius-flux trends. In H/He-dominated atmospheres, one of the most important atmospheric composition indicators is the C/O ratio. The C/O ratio plays a critical role in controlling the 
observable features, and has been hypothesized to be set by: 1) where and when a planet forms in the disk relative to ``snow lines'' of major C and O species, and 2) the relative accretion of gas vs.\ solids (see \citealt{Madhusudhan2016, Madhusudhan2019} and references therein). In smaller planets the C/O ratio can be influenced by many additional processes -- outgassing, vaporization, escape, impacts, photochemistry, weathering, and even biology \cite[e.g., ][]{Hu2012,Lammer2014,Gaillard&Scaillet2014,Schaefer&Fegley2017,He2018, Zahnle2020} -- let alone the composition of the initially-accreted solids \citep{elkins-tanton&seager2008, Schaefer&Fegley2010}. A given atmosphere may also change in oxidation state over time. Thus, while teasing out the implications of the atmospheric C/O ratio in smaller planet formation will likely be challenging, it is a natural place to begin investigating trends in composition versus radius and flux.
Given the significant theoretical and observational uncertainty in how we should expect atmospheric C/O to vary with radius or flux, we stress that our injected population model is not meant to represent a physically plausible model. Instead, this fiducial population relation is only used to determine how sample selection can affect retrieved inferences about transitions in atmospheric composition as a function of radius and flux. 
Furthermore, to ensure that our conclusions are robust to the random draws involved in our simulation study, we conduct 10 trials of each of the three planet samples. In Figure \ref{fig:flowchart} we show a visual overview of the steps described in the following subsections.

\subsection{Injecting a Fiducial Population Model}\label{subsec:fiducpopmod}

To begin our simulation study, we first define a simple C/O ratio based only on the abundances of three molecules:
\begin{equation}\label{eqn:ctoo}
C/O = \frac{\chi_\mathrm{CO_2}+\chi_\mathrm{CH_4}}{2*\chi_\mathrm{CO_2}+ \chi_\mathrm{H_2O}}
\end{equation}
where $\chi$ represents the abundance of each molecule. We focus specifically on these three molecules as they are the expected dominant sources of opacity in NIRSpec G395H's 3-5 $\mu$m region for the planets explored here (see Figure \ref{fig:samples}). Of course, there are other C and O bearing species that have the potential to affect the C/O ratio, and that would give valuable context clues to the nature of these planet atmospheres. For example, the vertically distributed abundance of CO (along with CH$_4$) is sensitive to many parameters such as vertical mixing, gravity, temperature, and metallicity \citep{Zahnle2014ch4co}. Additionally, CO along with HCN, C$_2$H$_2$, and C$_2$H$_6$ have been identified as molecules that could help distinguish the existence of a shallow  surface ($<$10~bar) typical of rocky planets, versus a deep surface ($>$100~bar) typical of gaseous planets \citep{Yu2021identify}. These molecules also have absorption bands in the 3-5~$\mu$m region. However, as we motivate in \S\ref{subsec:spectra}, including these physical processes would require complex atmospheric modeling that is beyond the scope of this analysis. Instead, we bypass complex 
modeling and directly choose how to vary C/O with respect to $R_p$ and $F_{\rm{insol}}$.  Our results are therefore based on the specific choices for this injected population relation, which we motivate below.

Next, we must choose how to vary C/O with respect to $R_p$ and $F_{\rm{insol}}$. 
From in-depth retrieval and information content studies that utilize simulated JWST data \citep{Greene2016JWST, BatalhaLine2017}, the expected precision obtained on log(${C/O}$) will likely be on the order of 0.5-1.5 dex, if not upper/lower limits. Simply put, for small planets we may only obtain order of magnitude constraints on C/O ratio. Therefore, only the most basic of population relations can be determined.  Given our primary goal of determining the feasibility of unearthing a possible trend with JWST-quality spectra and how that depends on the selected sample, we need a relation that can be informative even in the presence of large error bars and that allows for significant variability among the individual planets' C/O ratios.  One such population model is a logistic function that classifies a planet as having a carbon- or oxygen-dominated atmosphere ($\log{C/O}\ge0$ and $\log{C/O}<0$, respectively), where the atmospheric state depends on the $R_p$ and $F_{\rm{insol}}$ as
\begin{equation}\label{eq:poprelation}
\ln{\frac{P(\log{C/O}<0)}{P(\log{C/O}\ge0)}} = m_r R_{p} + m_f F_{\rm{insol}} + b 
\end{equation}
In this logistic relation, $P(\log{C/O}<0) \equiv P_O$ is the probability that the atmosphere is oxygen-dominated, $P(\log{C/O}\ge0)\equiv P_C$ is the probability that the atmosphere is carbon-dominated, and $m_r$, $m_f$, and $b$ are the population parameters that define how the relative likelihood of being carbon- or oxygen-dominated varies with respect to the planet's radius and insolation flux.  Note that the left-hand side of the relation is not C/O itself, but the \emph{probability} that a planet has a C/O ratio below 1 relative to the \emph{probability} that a planet has a C/O ratio above 1. By focusing on the probability that the target variable (C/O in our case) has a value in a certain range instead of focusing on the value of the variable itself, logistic relations define boundaries in the parameter space of the regressor variables ($R_{pl}$ and $F_{insol}$) that simultaneously enable classifications (carbon-dominated if $P_{O,i} < 0.5$ or oxygen-dominated if $P_{O,i} \ge 0.5$) and allow the target variable to take on a wide range of values across the population.  This parameterization therefore enables us to assess the presence of transitions in planetary atmospheres as a function of radius and insolation flux, even with uncertain atmospheric C/O measurements.

To use Equation \ref{eq:poprelation} in our simulation study, we must first specify values for the population parameters $m_r$, $m_f$, and $b$. To do this, note that Eq. \ref{eq:poprelation} describes a plane in the ($R_{pl}$,  $F_{insol}$, ln$(P_O/P_C)$) space; to define a specific plane, we choose three points that are reasonable based on our current understanding of exoplanet atmospheres and solve the resulting system of three equations for $m_r$, $m_f$, and $b$.  Specifically, these points are\footnote{Note that the above logistic relation does not allow a probability to be strictly zero, in which case the log ratio on the left-hand side would be undefined.}: $P_O=0.5$ at ($R_{pl} = 1 R_\oplus,F_{insol} = 1000 F_\oplus$), $P_O=0.01$ at ($R_{pl} = 3 R_\oplus,F_{insol} = 1000 F_\oplus$), and $P_O=0.99$ at ($R_{pl} = 1 R_\oplus,F_{insol} = 1 F_\oplus$), which gives $m_r=-2.30 $, $m_f=-1.53$, and $b=6.89$. For potentially rocky planets, this $P_O$ represents the hypothesis that those planet's atmospheres are more likely to be oxidized, similar to present day atmospheres of the terrestrial planets in our own Solar System. For example, CO$_2$, not  CH$_4$, is the major carbon bearing species in Venus, Earth, and Mars \citep{wayne1991chemistry}. For potentially gaseous planets, this  $P_O$ corresponds to the hypothesis that these planet's atmospheres are more likely to have near solar C/O ratios. 
However, we emphasize that these probabilities do not correspond to verified hypotheses in the study of exoplanet atmospheres. Additionally, we note that these probabilities test the optimistic case that $P_O$ actually has a transition with radius from low (near zero) to high (near 1) probability. We acknowledge that many other scenarios are possible, 
including the scenario where there is no such transition or a weak transition. However, choosing such cases would prevent us from addressing our goal of determining if population-level inferences can be made by studying a sample of planets with JWST. We view our choice in $P_O$ as a starting point.

From this fully specified population relation we then plug in each planet's $R_{pl}$ and $F_{insol}$ to obtain $P_{O,i}$ and $P_{C,i} = 1-P_{O,i}$.
Next, we draw the planet's $\log{C/O}$ from a uniform distribution with bounds of $(-2,0)$ or $(0,2)$, where the $<0$ or the $>0$ range is chosen in proportion to the drawn value of $P_{O,i}$.  With a planet's specific C/O value in hand, next we must use it to determine the individual molecular ratios CO$_2$/H$_2$O, CH$_4$/H$_2$O, and CO$_2$/CH$_4$.  To do this uniquely, we must independently specify at least one of these three molecular ratios; we choose CO$_2$/CH$_4$, which we draw from a uniform distribution. We specifically do not choose to draw atmospheric ratios using the assumption of chemical equilibrium since this physical condition will not apply to the full sample of planets here (namely those that are potentially rocky, and/or cool (T$_\mathrm{eff}<1000$~K).  To determine the bounds of this distribution, we first consider that the ability to detect a species in transmission can be roughly determined by assessing the ratio of the cross sections, $\sigma$, weighted by the molecular abundance, $\xi$. At a given wavelength, $\lambda$, if $
   \sigma_{\lambda,CO_2} \xi_{CO_2} >> \sigma_{\lambda_,CH_4} \xi_{CH_4}
$, CO$_2$ will dominate the spectrum. Across 3-5~$\mu$m, the median ratio of the cross sections is $\sigma_\lambda{CO_2}$/$\sigma_\lambda{CH_4}\sim$10 (similarly, $\sigma_\lambda{CH_4}$/$\sigma_\lambda{H_2O}\sim$15). Therefore, we choose $-2 <$log(CO$_2$/CH$_4$)$<0$, which creates a diverse set of resultant spectra spanning cases with detections of both carbon-bearing species to detections of only a single carbon-bearing species. 

With this population relation defined, and C/O and CO$_2$/CH$_4$ values drawn for each planet, we have begun to specify the atmospheric state of each planet, which we continue in \S \ref{subsec:spectra}.  

\subsection{Atmospheric Modeling Parameters} \label{subsec:spectra}
We choose a simple methodology for simulating observed planetary spectra, rather than creating ``self-consistent'' models (relying on converging temperature, chemical models, and cloud profiles based on initial boundary conditions). This is in line with many theoretical studies that have sought to determine the detectability of super-Earth and sub-Neptune atmospheres in the JWST era \citep{Morley2017,BatalhaLine2017, Batalha2018,Chouqar2020,G-Mesa2020}.  
For example, \citet{Morley2017} created a grid of models with Earth-, Titan-, and Venus-like atmospheres, and \citet{Batalha2018} chose a fixed background gas scenario (e.g. H$_2$O-rich vs H$_2$-rich) and varied the remaining trace species in fixed increments. These modeling choices are especially necessary for planets that straddle the super-Earth/sub-Neptune regime, because mass/radius cannot serve as a reliable proxy for H/He envelope mass below about 2.2 R$_{\oplus}$, at which point internal structure models with significant fractions of heavier gases like H$_2$O are also able to fit the observed exoplanet masses and radii \citep{Valencia2006,rogers2015}. 

Here we choose a modeling framework that affords us the opportunity to address the main goal of this work -- the importance of sample selection. 
We use a double-grey analytical parameterization for the temperature-pressure profile \citep{guillot2010}. We only consider 4-molecule atmospheres (H$_2$, H$_2$O, CO$_2$, and CH$_4$). Similar simplicity has been given to other studies investigating hypothetical atmospheres (e.g., 100\% H$_2$O in \citet{Greene2016JWST}; H$_2$-H$_2$O in \citet{Batalha2017challenges}). We assume each atmospheric composition to be well-mixed (i.e. uniform with altitude). These well-mixed values are chosen based on the injected population relation. 

We note that opting for this simplicity ignores two important factors in determining population-level trends that would exist in nature. First, our simplicity  may exclude physical processes that could potentially provide vital context clues regarding the nature of super-Earth/sub-Neptunes. 
For example, \citet{Yu2021identify} identified seven chemical species that could help distinguish the existence of  shallow versus deep surfaces.  Second, our simplicity may exclude physical processes that would make it more difficult to establish trends in atmospheric parameters. For example, the stellar UV flux from each parent star determines to what degree photochemical processes will drive chemical abundances of key molecules such as CH$_4$ \citep[e.g.,][]{Hu2021Photochemistry}. Both processes analyzed in \citet{Yu2021identify} and \citet{Hu2021Photochemistry} require robust photochemical modeling with complete chemical networks. Therefore, despite these likely effects, incorporating them in a uniform manner is not trivial and beyond the scope of this analysis.  

Given the chosen model, and with the chosen C/O, CO$_2$/CH$_4$, and H$_2$O/CH$_4$ uniquely assigned (see \S \ref{subsec:fiducpopmod}), the final parameters to choose are bulk H$_2$ fraction and cloud parameters. Higher bulk H$_2$ fraction increases the scale height of the atmosphere because of a decreased mean molecular weight, and thus increases the magnitude of spectral features. Increased cloud optical depth/decreased cloud pressure has the effect of muting spectral features. For JWST-quality data, the effect of these two parameters on the spectra are degenerate in the infrared ($\gtrapprox2\mu$m) \citep{Benneke2012, Batalha2017challenges}. Additionally, even for hot Jupiters it is not yet clear 
how to predict the degree of cloud coverage as a function of planet parameters (e.g., equilibrium temperature and gravity) \citep{Wakeford2019,Gao2020,Alam2020pancet}. The same is true for super-Earths/sup-Neptunes, which are undoubtedly more difficult targets for high SNR spectra \citep{Crossfield2017trends,Dymont2021cleaning}. 

Therefore, we start by exploring the cloud-free case with a bulk H$_2$ fraction fixed to 99\%. This  H$_2$ value is chosen to correspond to a Neptune-like H$_2$/He-fraction (100$\times$Solar metallicity, \citet{karkoschka1998methane}). For completeness, we also compute spectra with a bulk H$_2$ fraction of 90\%  as well as a ``cloudy'' case. In our ``cloudy'' scenario we insert a grey opacity source at 0.1 bars -- a case where the observation is limited by the tropopause of the planet, which is defined at 0.1 bars in all Solar System planets \citep{Robinson2014}. We find that if we prescribe a lower H$_2$ bulk fraction and included the effect of clouds as grey opacity source at a fixed pressure, there are 4-5 planets in each sample with no detectable features. Therefore, all three samples will be similarly encumbered by clouds and increased mean molecular weight via an effective decrease in sample size. We discuss this limitation in our concluding remarks. 

\subsection{Modeling Spectra and Retrieving Abundances} \label{subsec:retrievals}
With the given prescription for atmospheric abundances and temperature-pressure profiles, we use the \texttt{PICASO} radiative transfer tool \citep{picaso2019,picaso2020} to compute the transmission spectra. Of importance for this analysis are the opacities of H$_2$O, CH$_4$ and CO$_2$, for which we use \citet{Polyansky2018-Water-POKAZATEL,Yurchenko14, Huang14}, respectively. Our $R=10^6$ line-by-line opacities are resampled at $R=10^4$ to be suitable for retrievals at $R=100$ and are available for download at \citet{picaso_db2020}. For each transmission spectrum, we use \texttt{PandExo} to compute a simulated JWST observation with NIRSpec G395H, which we then bin to $R$=100 for the ultimate retrieval. Lastly, 
we pair \texttt{PICASO} with the open source Nested Sampler \texttt{dynesty} \citep{dynesty}, which implements the algorithm developed by \citet{Skilling2004Bayes}. \citet{mukherjee2021} outlines the specific hyper-parameters used for the sampler, \texttt{dynesty}. 

For each planet we compute the posterior probability distributions for five free parameters: 1) the irradiation temperature of the \citet{guillot2010}-P(T) profile (in K; prior: U(300,1200)); 2) H$_2$ bulk abundance (in dex; prior: U(-6,0)); 3) H$_2$O/CH$_4$ abundance ratio (in dex; prior: U(-6,6)); 4) CO$_2$/CH$_4$ abundance ratio (in dex; prior: U(-6,6)); and 5) $x$R$_p$, a scaling factor to the reported radius derived from the \textit{Kepler}/TESS transit observation that we arbitrarily define at 10 bars (unitless; prior: U(0.5,1.5)). 
For this analysis we focus specifically on the retrieval results of the abundance ratios and combine the posteriors for H$_2$O/CH$_4$ and CO$_2$/CH$_4$ following Eq. \ref{eqn:ctoo} to compute a posterior for C/O. In total we run 23 unique planetary atmospheres (some of the selected targets are the same across the three samples) for each of the 10 random trials, for a total of 230 retrievals.

\section{Statistical Model for the Population} \label{sec:hbm}

In \S \ref{subsec:fiducpopmod} we describe the population relation between planetary atmospheric C/O, radius, and insolation flux that anchored our simulations of atmospheric spectra.  To assess how well we can recover that population relation using each of the three samples outlined in \S \ref{sec:targets}, we must define a statistical model that will enable us to: 1) infer the parameters of that relation from the individual atmospheric retrievals; 2) quantify the uncertainty in those inferred parameter values; 3) and compare the inferred values to the ``true'', injected values.  Specifically, we are interested in how well we can recover the values for $m_r$, $m_f$, and $b$ in Eq. \ref{eq:poprelation}; respectively, these hyperparameters control the steepness of the transition from log(C/O) $<0$ to log(C/O) $>0$ as a function of planet radius, control the steepness of the transition as a function of insolation flux, and set the constant probability that any given planet in our sample has an atmosphere with log(C/O) $<0$, regardless of radius or insolation flux.

To infer these three hyperparameters from the individual planets' C/O posteriors, we use the following hierarchical Bayesian model:
\begin{align}\label{eq:HBM}
\pi(m_r)&=\text{U}(-10,10)\nonumber\\
\pi(m_f)&=\text{U}(-10,10)\nonumber\\
\pi(b)&=\text{U}(-15,15)\nonumber\\
\ln{\Big(\frac{P_{O,i}}{1-P_{O,i}}\Big)} &= m_rR_i + m_f\log(F_i) + b\nonumber\\
\mathcal{L}(\bm{X}|\bm{P_O},m_r,m_f,b)&=\prod^{N}_{i=1}P_{O,i}X_i + (1-P_{O,i})(1-X_i)  
\end{align}
Recall that $P_{O,i}$ is the probability that the i-th planet's true C/O ratio $<1$ (and so $P_{C,i} = 1-P_{O,i}$ is the probability that its true C/O ratio $\ge 1$).  Additionally, $X_i$ is the fraction of a planet's retrieved C/O posterior that lies below C/O$=1$: $X_i = \int^1_{-\infty}\mathcal{P}_i(C/O)d_i(C/O)$, where $\mathcal{P}_i(C/O)$ is the posterior probability of the i-th planet's C/O ratio, i.e. the ``retrieved'' C/O distribution that is outputted from the individual atmospheric spectral analyses (see \S \ref{subsec:retrievals}).

In the above hierarchical model we first specify the prior distributions on the hyperparameters, denoted with $\pi()$, to be uniform on a range that spans very steep transitions ($m=10$) to no transition ($m=0$) to very steep transitions in the opposite direction ($m=-10$). However, in a case such as the one presented here -- a relatively small sample and sometimes weak constraints on individual-planet parameters (\S5) -- it is necessary to ensure the posteriors are not dominated by the choice of priors. Another principled choice is to choose a uniform prior on $\arctan{m}$ in order to sample the angle of the plane created by $m_r$ and $m_f$. In this way we place uniform priors on the angle described by this slope, not the slope itself while keeping the prior on $b$ unchanged. The comparison and resulting implications are discussed in \S \ref{sec:results}.

The prior distribution on $b$, which represents a constant log-odds that the planet is oxygen-dominated, allows it to span essentially 0 to essentially 1 (within a factor of $e^{-15}$; note the natural logarithm on the left-hand side of the fourth line of Eq. \ref{eq:HBM}).  The logistic relation of Eq. \ref{eq:poprelation} follows, which gives the log-odds that a planet's true C/O is less than 1, is based on its radius and insolation flux and the (free to vary) hyperparameter values.  

Lastly, the model contains the likelihood that the retrieved C/O ratios follow the provided logistic relation. 
Note that this hierarchical model does not include measurement uncertainty in the planet radii ($R_i$) or the insolation fluxes ($F_i$). For planets transiting bright, well-studied stars, the uncertainties in these planet parameters will be much smaller than the uncertainties on the C/O ratios; to keep the model as simple as possible, we do not include these measurement uncertainties.  

To evaluate this hierarchical Bayesian model, we use pyStan, a Python interface for Stan \citep{Stan2019}, a probabilistic coding language which allows users to directly specify the likelihood and prior distributions of a Bayesian model and which performs Markov Chain Monte Carlo (MCMC) sampling of the resulting posterior to enable parameter estimation.  The particular MCMC algorithm implemented by Stan is Hamiltonian Monte Carlo with a No U-turn Sampler, which probes the ``potential energy'' contours of the posterior probability distribution with trajectories in parameter space that have ``momentum'' from one step in the Markov chain to the next.  To estimate $m_r, m_f$, and $b$ of Eq. \ref{eq:HBM}, we computed 8 Markov chains of 100,000 steps each.  Dropping the first half of each chain for burn-in and thinning the resulting samples by a factor of 25, we retained a total of 16,000 MCMC samples.  MCMC performance metrics like $\hat{R}$ and the number of effective samples indicate excellent convergence ($\hat{R} = 1.0$ and $n_{eff} \sim 16000$ for all parameters).  To compute the hyperparameter values we report in \S \ref{sec:results}, we perform a three-dimensional kernel density estimate of the saved MCMC samples and take the mode of that distribution.

\begin{figure*}[htbp]
\centering
\includegraphics[width=\textwidth]{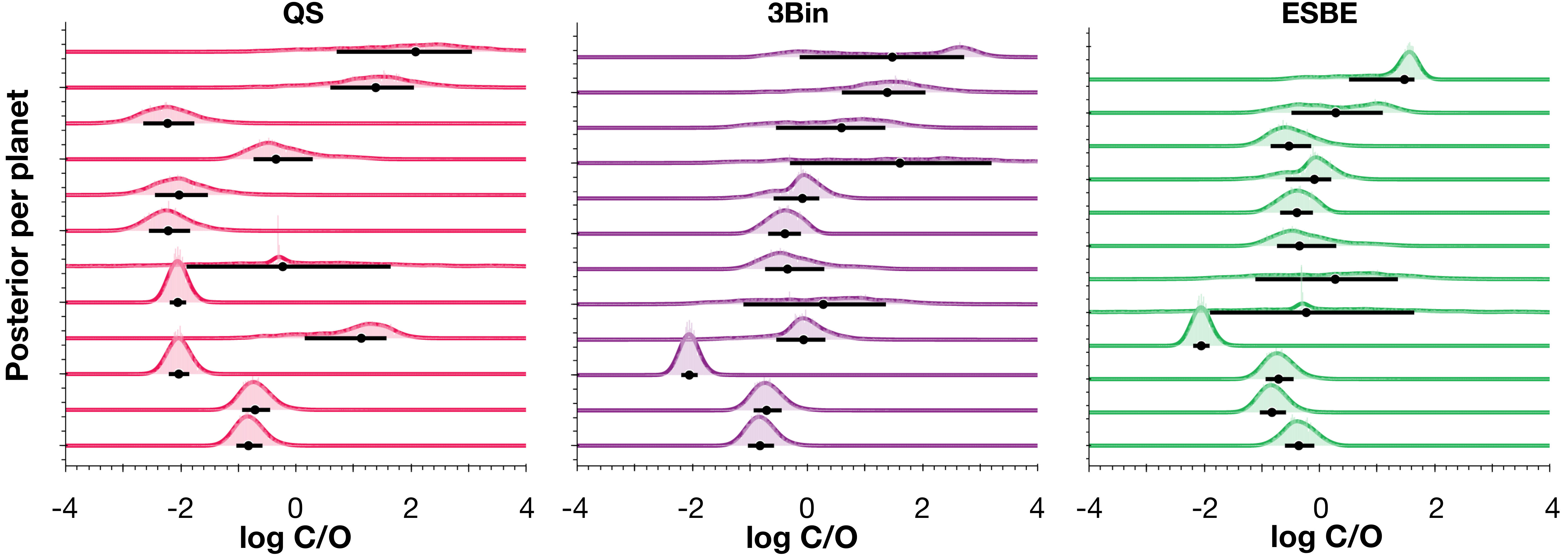} 
\caption{The posterior probability distributions for individual-planet atmospheric C/O (see Eq. \ref{eqn:ctoo} for C/O definition) for a representative trial of all three simulated surveys (Trial 0, for reference). In each plot, posterior distributions for each planet are ordered from small planet radii (bottom of plot) to large (top). Black lines indicate the 1$\sigma$ credible interval for each planet's log C/O. Note, in some cases the CI is large due to a degenerate log C/O solution (e.g. the top double-peaked posterior in 3Bin sample). For each of the three samples (QS, ESBE, and 3Bin), the expected log C/O constraints range from 0.1-2 dex across all planets, consistent with previous explorations of JWST capabilities.} 
\label{fig:atmopost}
\end{figure*}

\begin{figure*}[htbp]
\centering
\includegraphics[width=\textwidth]{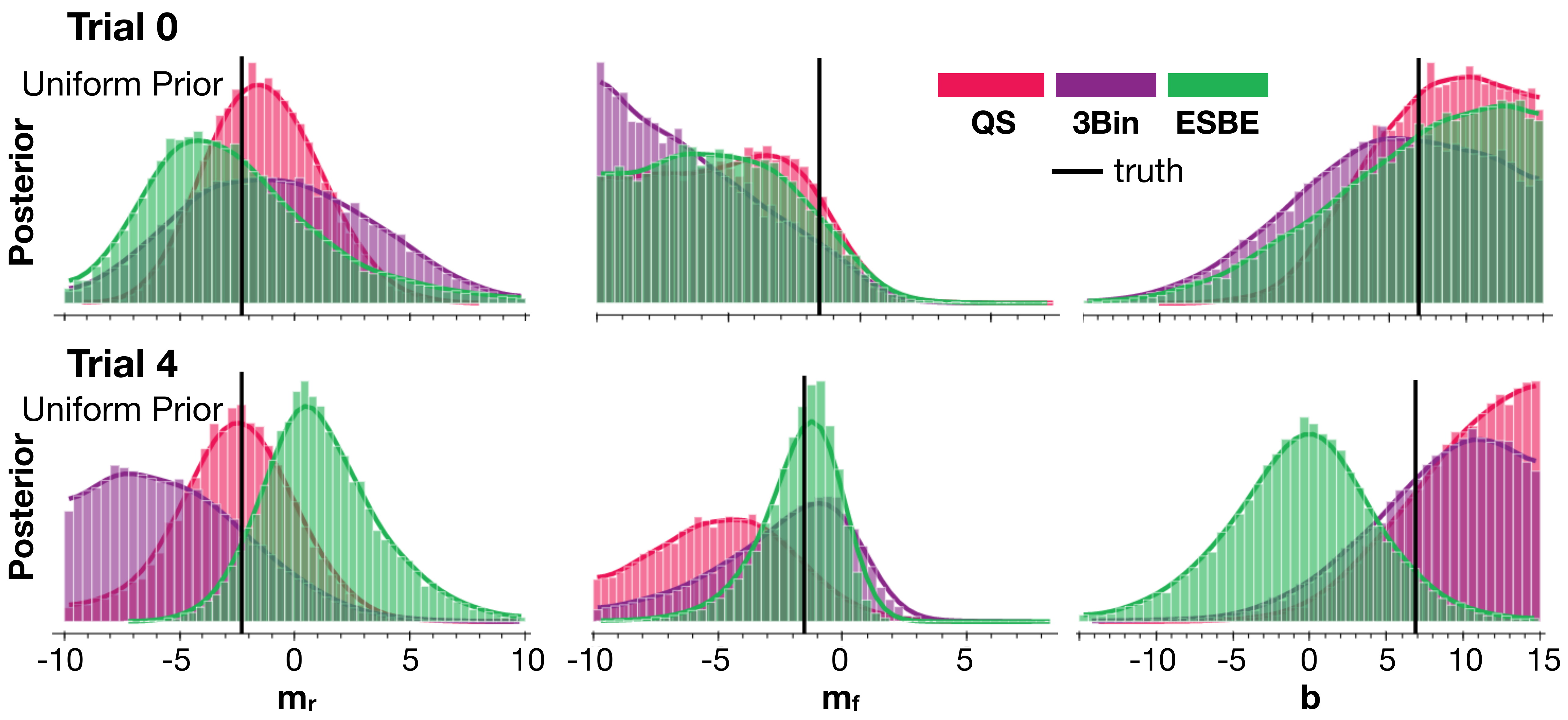} 
\caption{The posterior probability distributions for the three population parameters (see Eq. \ref{eq:poprelation}) for two representative trials (Trial 0, and 4) and a uniform prior on the population-level parameters. True injected values are shown with black vertical lines. Credible intervals on population parameters differ drastically between different hypothetical surveys (i.e. QS, ESBE, and 3Bin).} 
\label{fig:poppost}
\end{figure*}

\begin{figure}[htbp]
\centering
\includegraphics[width=\columnwidth]{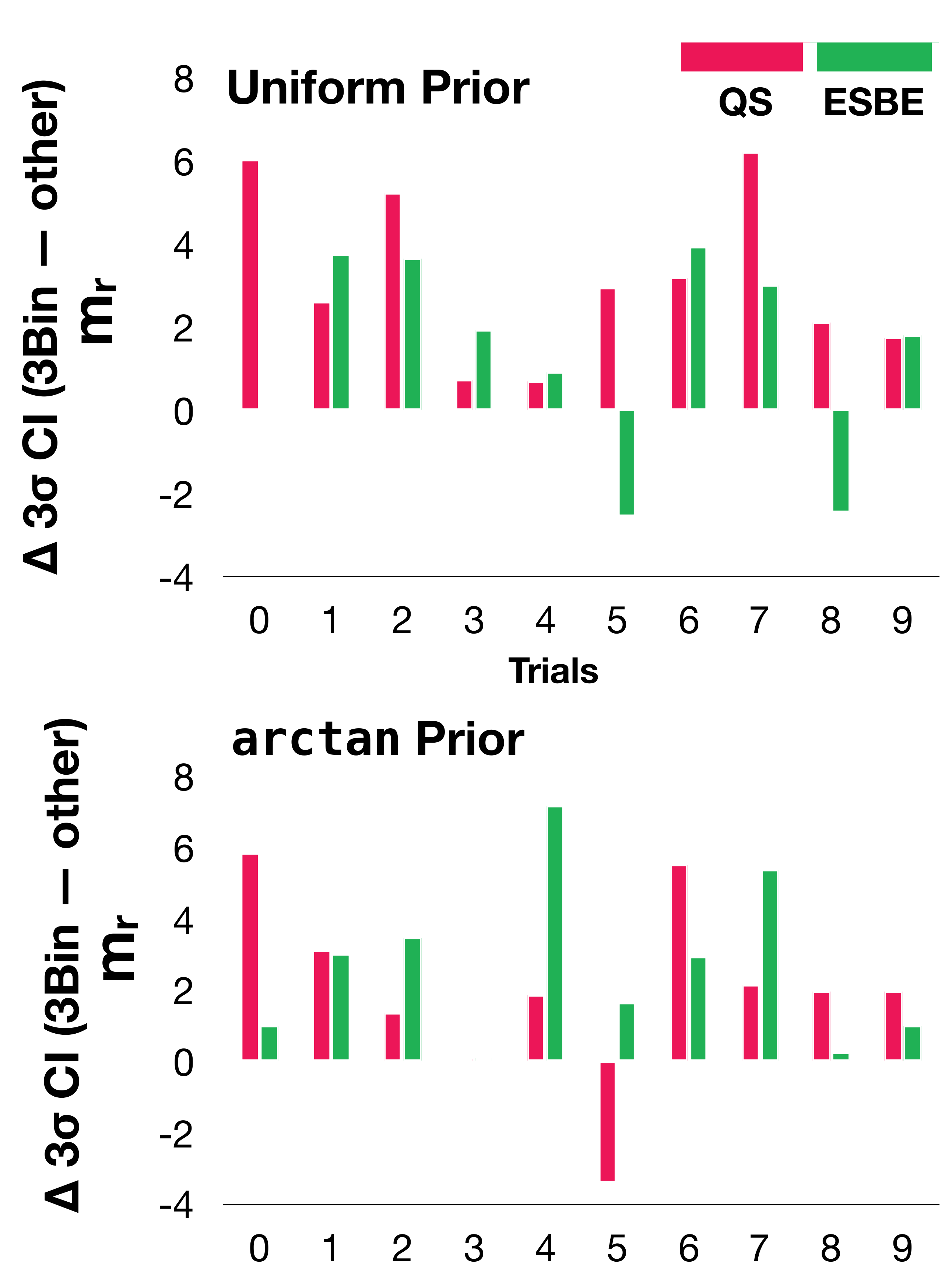} 
\caption{Differences between 3$\sigma$-credible intervals for the population parameter describing the radius relation, $m_r$, for the 3Bin sampling method versus the QS and ESBE sampling methods, with two choices of prior. As shown in Figure \ref{fig:poppost}, $m_r$ is the only parameter with Gaussian credible intervals, as opposed to upper or lower limits on the other parameters. The QS and ESBE sampling methods result in smaller credible intervals on m$_r$ for nearly all trials when compared to the 3Bin method. } 
\label{fig:delta}
\end{figure}

\begin{figure*}[htbp]
\centering
\includegraphics[width=\textwidth]{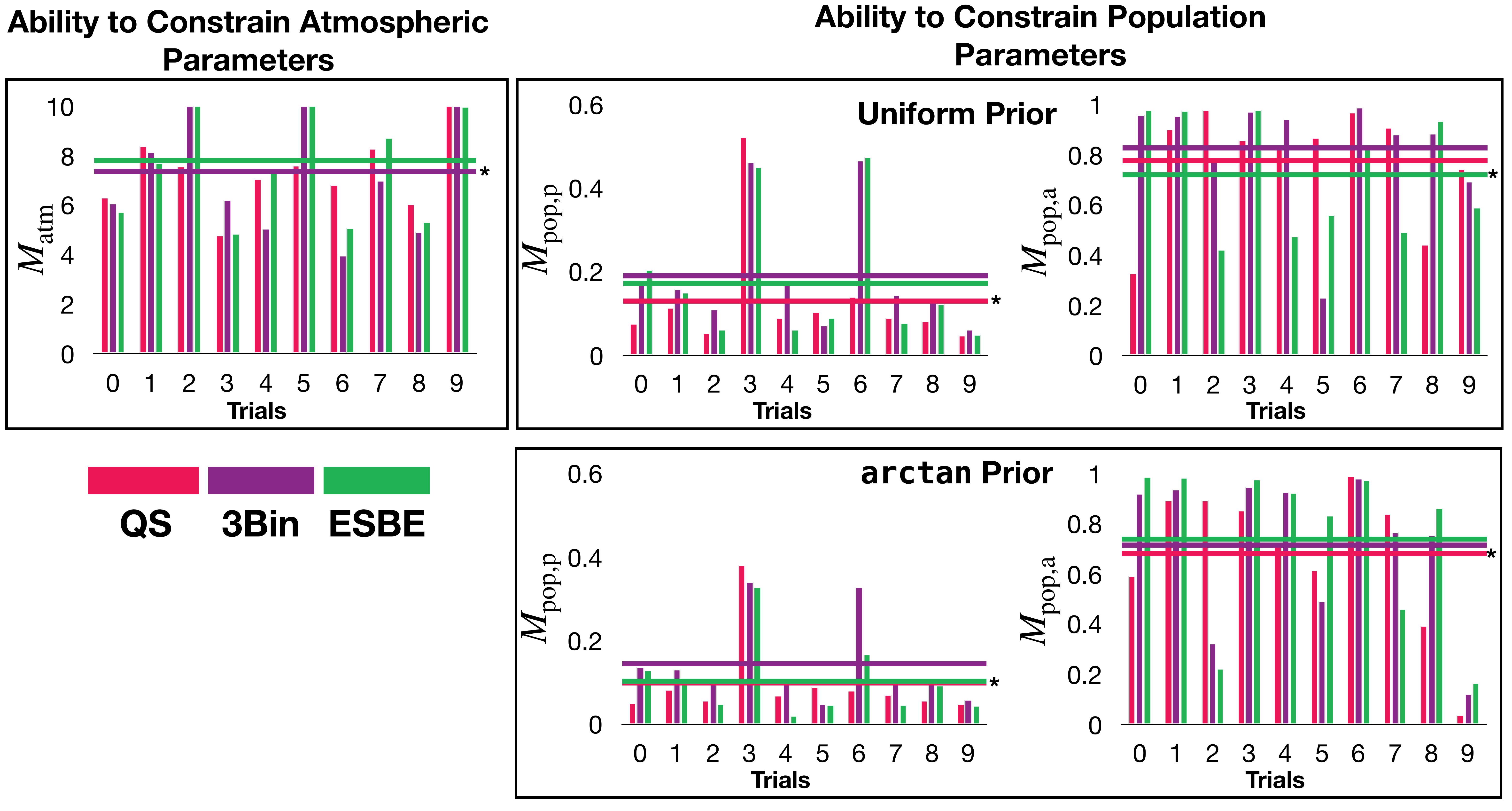} 
\caption{The result of three metrics used to evaluate the success of the three hypothetical NIRSpec G395H surveys (QS, ESBE and 3Bin). The first metric (left panel) describes the accuracy and precision of the constrained atmospheric C/O ratio (shown for individual planets in Figure \ref{fig:atmopost} for Trial 0). The second and third metrics (middle/right panel) describe the ability to constrain the population parameters precisely and accurately, respectively. For all three metrics, the lower the value, the ``better'' the survey performed (indicated by \text{*}). Horizontal lines indicate the mean across all 10 trials (for $M_\mathrm{atm}$ the purple and pink lines are overlapping). Overall, the QS- and 3Bin- derived samples have the highest success in constraining atmospheric parameters, while the QS- and ESBE- have the highest success in constraining population parameters.} 
\label{fig:sumstats}
\end{figure*}

\section{Results}\label{sec:results}

We performed ten trials of the analysis depicted in Figure \ref{fig:flowchart} to understand the overall performance of each sample selection method. For each trial, there is randomness when drawing each planet's $\log$(C/O) and CO$_2$/CH$_4$ abundance ratios from a uniform distribution (see \S 3.1). 
First, we highlight the results of representative trials to showcase the typical behavior of 
retrieving the individual planets' atmospheric C/O ratios and the three population parameters drawn for each trial. Then, we discuss the success of the three simulated surveys across all 10 trials. Lastly, we test the result's dependence on the choice of population-level prior (either uniform or \texttt{arctan}).  

Figure \ref{fig:atmopost} shows the individual posteriors of the derived atmospheric C/O ratio for one trial of each of the three 12-planet surveys (Trial 0). The range in retrieved 1$\sigma$ constraint intervals for log C/O span $\pm$ 0.1-2 dex for all three simulated surveys. This range remains consistent across all 10 trials. This large span in log(C/O) ratio is primarily due to the random draws in CO$_2$/CH$_4$. For example, TOI 836.01  obtained a 1$\sigma$ log(C/O) constraint of $\pm$1.4 in Trial 0 (see top most posterior in QS Figure \ref{fig:atmopost}). In Trial 4 the 1$\sigma$ interval was nearly a third of that value,  $\pm$0.5. In Trial 0 the random draw of C/O and CO$_2$/CH$_4$ resulted in log(H$_2$O/CH$_4$) and log(CO$_2$/CH$_4$) of -1.96 and -2.84, respectively. In Trial 4, the same values were -0.43 and -0.37, respectively. As noted in \S3.1, when the relative abundances of log(H$_2$O/CH$_4$) and log(CO$_2$/CH$_4$) are $>> or << 1$, the 
molecule in higher abundance will 
dominate the spectrum, leading to only one 
molecular detection. Therefore, the random draw of Trial 0 makes it more difficult to constrain log(C/O) because only CH$_4$ is detectable in the spectrum. Overall, the span of the 1$\sigma$ log(C/O) credible intervals are consistent with previous investigations of the capabilities of JWST \citep[e.g.][]{Greene2016JWST}. We provide a detailed comparison between each sample in \S\ref{sec:metric}.


Figure \ref{fig:poppost} shows the marginalized posterior distributions on the three injected population hyperparameters 
(see Eq. \ref{eq:poprelation}) given the atmospheric constraints on log C/O for two representative trials (Trials 0 and 4) with a uniform prior. Note that the marginalized posterior distributions are visualized in one dimension but are in fact three-dimensional volumes. Two immediate results can be drawn from Figure \ref{fig:poppost}. The first is that the posteriors for $m_f$, $m_r$, and $b$ are different for each trial. The second is that for population parameters $m_f$ and $b$, in some cases only upper/lower limits are achieved, as opposed to Gaussian constraints. It is worth emphasizing that 
an upper/lower limit on a population parameter would still be novel and valuable scientific insight. 
For example, in Trial 0 for the insolation flux population parameter, $m_f$, all three cases suggest that it is negative, which would mean that the fraction of planets with oxygen-dominated atmospheres decreases as insolation flux increases. 

Further comparing the two trials shown in Figure \ref{fig:poppost} with the same uniform prior assumptions on population-level parameters, we see that in the example of Trial 0 (upper panel), all three hypothetical surveys can produce an upper limit on $m_f$ and $b$. For $m_r$, the QS-derived sample results in a relatively precise and accurate constraint of $m_r=-1.4_{-2.3}^{+2.5}$ compared with, for example, 3Bin's $m_r=-0.9_{-4.0}^{+4.4}$.  In the example of Trial 4, the ESBE-derived sample produced both the most precise and the most accurate credible intervals on parameter $m_f$. 

Overall, we find that across the three sampling methods and trials, we are able to retrieve a Gaussian constraint on the radius population parameter, $m_r$. This allows us to directly compare their credible intervals across trials. The results of a comparison between the 3$\sigma$ credible intervals derived from the 3Bin sampling method versus the QS and ESBE sampling methods are shown in Figure \ref{fig:delta} for both assumptions of population-level prior. For the choice of uniform prior across all trials, the 3Bin sample produces larger $3\sigma$ credible intervals on the radius population parameter, $m_r$, as compared with the QS sample. It also produces larger $3\sigma$ credible interval on all but two trials, when compared to the ESBE sample. A similar result is obtained for the choice of \texttt{arctan} prior (Figure \ref{fig:delta} bottom panel). Although this provides some evidence that QS and ESBE outperform the 3Bin method, there are other factors to consider. Specifically, we introduce metrics in the following \S\ref{sec:metric} that consider both precision and accuracy across all population parameters.


\subsection{Three Metrics for Survey Success Evaluation}\label{sec:metric}
We consider three metrics to evaluate how each hypothetical NIRSpec G395H survey did across all ten trials, and whether there is a clear best-practice for constructing atmospheric surveys. 

The first metric, $M_\text{atm}$, describes the overall precision and accuracy of the individual-planet atmospheric parameter constraints, which in this study is the posterior probability distribution of C/O. 
This quantity is computed by combining the retrieved posterior distributions of H$_2$O/CH$_4$ and CO$_2$/CH$_4$ via the C/O approximation defined in Eq. \ref{eqn:ctoo}. For each trial 
we compute the chi-squared of the injected ($C/O_{inj,i}$) vs. retrieved ($C/O_{ret,i}$) log C/O ratio over all 12 planets, while accounting for the 1$\sigma$-posterior width of each planet ($\sigma_{i}$): 
\begin{equation}
    M_\text{atm} = \sum_{i=1}^{12} \frac{(C/O_{ret,i} - C/O_{inj,i})^2}{\sigma_{i}^2}
\end{equation}
Therefore, the lower the $M_\text{atm}$, the better the hypothetical survey did overall in constraining individual-planet atmospheric composition parameters. 

The second metric, $M_{\text{pop,p}}$, describes the overall ability to constrain the population parameters precisely. For this metric, we compute the 
volume enclosed in the joint, 3-dimensional posterior of the three population parameters ($m_f$, $m_r$, and $b$) at 3$\sigma$. We also explored using the 1$\sigma$ or 2$\sigma$ posterior volume as the metric; the overall conclusions were unchanged. Similar to $M_\text{atm}$, the lower the $M_{\text{pop,p}}$, the tighter the width of the joint posterior and the better the survey did at precisely constraining the population parameters. 

The third metric, $M_{\text{pop,a}}$ describes the overall ability to constrain the population parameters accurately. For this metric, we compute the effective ``distance’’ between the posterior mode and the injected value. We define ``distance’’ as the  difference between the injected value and the posterior mode in the three-dimensional hyperparameter space. Similar to previous metrics, the lower the $M_{\text{pop,a}}$, the higher the accuracy and the better the survey did at accurately constraining the population parameters. 


Figure \ref{fig:sumstats} shows the results of the three metrics for all 10 trials and both prior assumptions on population-level parameters. With regard to the atmospheric parameters, 
QS and 3Bin each had 4 out of 10 trials with the lowest value of $M_\text{atm}$. In the other two trials, 
ESBE had the lowest value of $M_\text{atm}$. According to the mean $M_\text{atm}$ across all ten samples, the QS- and 3Bin- derived samples performed nearly equally (average lines are overlapping in Figure \ref{fig:sumstats} with $M_\text{atm}$=7.1).

With regard to the population-level metrics, considering uniform priors on $m_r$ and $m_f$, QS had the lowest value of $M_\text{pop,p}$ in 6 out of 10 trials, and the lowest value of $M_\text{pop,a}$ in 4 out of 10 trials. The ESBE survey had the lowest value of $M_\text{pop,p}$ in 3 out of 10 trials, and the lowest value of $M_\text{pop,a}$ in 5 out of 10 trials. This suggests that QS and ESBE performed more similarly while 3Bin consistently had the poorest performance.  We note that the variability across trials is quite high and that an estimate of the metrics' variances given by the 20th-80th quantile range (approximately the quantiles that would correspond a 1-$\sigma$ uncertainty interval for a normally distributed metric, which these are not) would yield overlapping bands.

Furthermore, we test the robustness of these results against our assumptions of priors on the population-level parameters, $m_r$ and $m_f$, as described in \S \ref{sec:hbm}. Considering the \texttt{arctan} priors on $m_r$ and $m_f$, the QS sample achieves the lowest $M_\text{pop,p}$value in 4 out of 10 trials while ESBE achieves the lowest value in 6 out of 10 (for the uniform prior it was 6/10, and 4/10, respectively). For the $M_\text{pop,a}$ metric, QS achieves the lowest value 6 out of 10, ESBE achieves the lowest value 3 out of 10, and 3Bin achieves the lowest value once (for the uniform prior it was 4/10, 5/10, and 1/10, respectively). Though the overall conclusions drawn from the metrics are not changed, the choice of prior does affect the results on a trial-by-trial basis. 

In Table \ref{tab:wins} we show a break down of which sample achieved the lowest metric values for each trial and prior choice. In 3 out of 10 trials, the sample with the lowest $M_\text{pop,p}$ metric was changed, and in 2 out of 10 trials the sample with the lowest $M_\text{pop,a}$ metric was changed (shown as bolded text in Table \ref{tab:wins}). In all cases, the \texttt{arctan} prior changed the ``winning'' sample to either ESBE or QS, not 3Bin. In other words, our overall result that the 3Bin-derived sample achieves the poorest performance is not prior dependent. We verify this by showing in Figure \ref{fig:poppost} the representative posteriors of Trial 9, where the choice of prior did affect the population-level metric but, qualitatively, the posterior distributions of the population parameters are not drastically affected.

\begin{table}[]
\begin{tabular}{ll|ll|ll|}
                            &                                & \multicolumn{2}{l|}{Uniform Prior}  & \multicolumn{2}{l|}{\texttt{arctan} Prior}   \\ \hline
\multicolumn{1}{|l|}{Trial} & $M\text{atm}$ & $M_\text{pop,p}$ & $M_\text{pop,a}$ & $M_\text{pop,p}$ & $M_\text{pop,a}$ \\ \hline
\multicolumn{1}{|l|}{0}     & ESBE                           & QS               & QS               & QS               & QS               \\
\multicolumn{1}{|l|}{1}     & ESBE                           & QS               & QS               & QS               & QS               \\
\multicolumn{1}{|l|}{2}     & QS                             & \textbf{QS}      & ESBE             & \textbf{ESBE}    & ESBE             \\
\multicolumn{1}{|l|}{3}     & QS                             & ESBE             & QS               & ESBE             & QS               \\
\multicolumn{1}{|l|}{4}     & 3Bin                           & ESBE             & \textbf{ESBE}    & ESBE             & \textbf{QS}      \\
\multicolumn{1}{|l|}{5}     & QS                             & \textbf{3Bin}    & 3Bin             & \textbf{ESBE}    & 3Bin             \\
\multicolumn{1}{|l|}{6}     & 3Bin                           & QS               & ESBE             & QS               & ESBE             \\
\multicolumn{1}{|l|}{7}     & 3Bin                           & ESBE             & ESBE             & ESBE             & ESBE             \\
\multicolumn{1}{|l|}{8}     & 3Bin                           & QS               & QS               & QS               & QS               \\
\multicolumn{1}{|l|}{9}     & QS                             & \textbf{QS}      & \textbf{ESBE}    & \textbf{ESBE}    & \textbf{QS}   \\ \hline           
\end{tabular}
\caption{\label{tab:wins} Metric results for each trial and each choice of prior on population level parameters, m$_r$ and m$_f$.  Bolded text emphasizes the cases for which the choice of prior affected the metric results.}
\end{table}

\begin{figure*}[htbp]
\centering
\includegraphics[width=\textwidth]{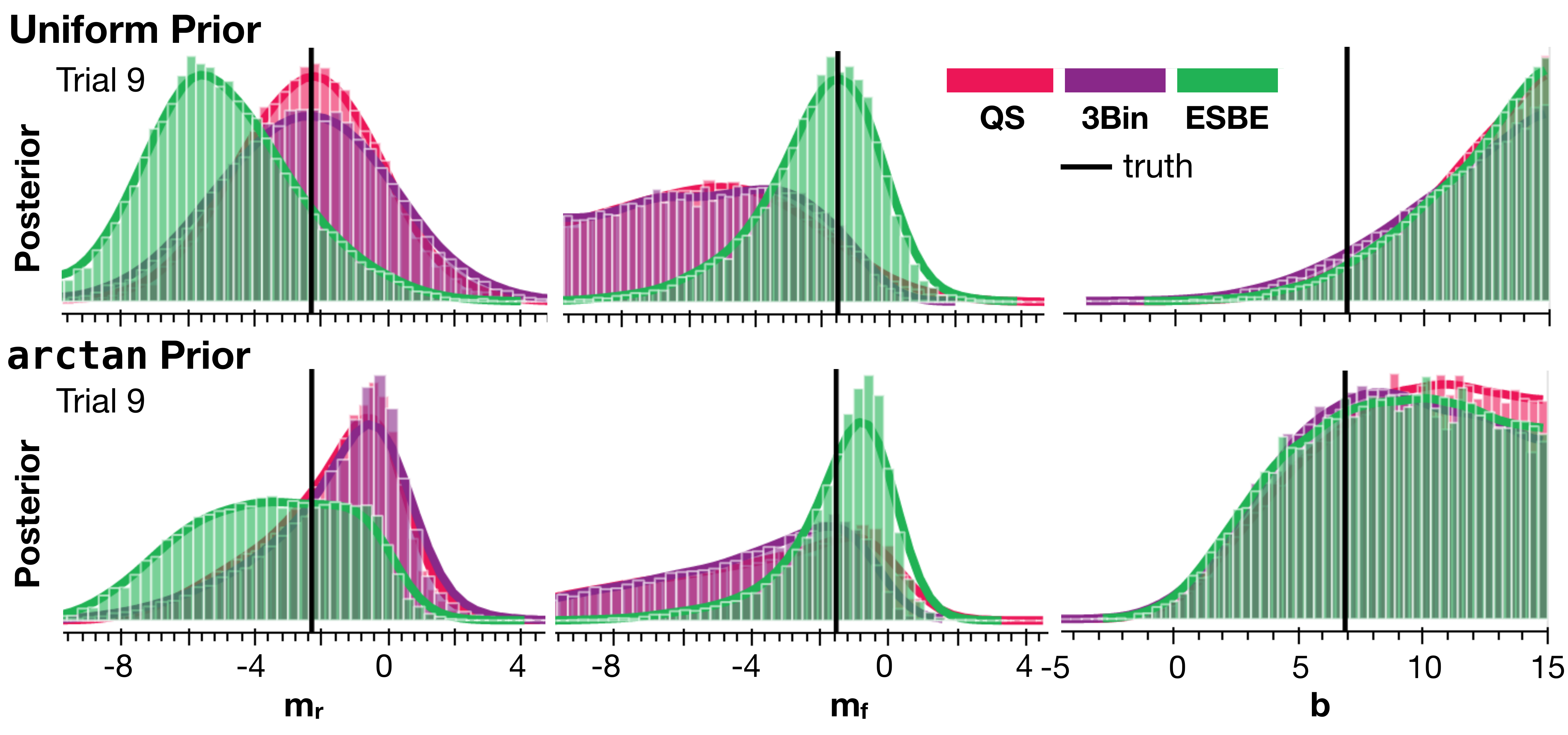} 
\caption{The posterior probability distributions for the three population parameters (see Eq. \ref{eq:poprelation}) for one trial and two different assumptions for prior on the population-level parameters. True injected values are shown with black vertical lines.} 
\label{fig:poppost}
\end{figure*}

Ultimately, 3Bin consistently had the poorest population parameter constraints, while QS and ESBE were relatively tied. 
According to the mean across 10 trials, with uniform priors on  $m_r$ and $m_f$, QS obtained slightly higher precision constraints, while ESBE obtained slightly higher accuracy constraints, and with \texttt{arctan} priors, QS obtained higher precision and accuracy constraints. However, we reiterate that a mean for the population parameters does not capture the variability across trials, which is significant as is evident from Figure \ref{fig:sumstats}. The variability exhibited from trial to trial with these 12-planet samples illuminates the critical need for samples larger than what can be feasibly done in a single JWST cycle. 

Ultimately, we consider QS and ESBE to have done equally well according to the number of times these samples achieved the lowest value of  $M_\text{pop,p}$ or $M_\text{pop,a}$. This particular result leads to an important conclusion: Even when the 3Bin sample had the ``best'' precision and accuracy on log(C/O), compared to the QS- and ESBE- samples, 3Bin resulted in population parameters that were worse in terms of both accuracy and precision. This finding showcases why sample selection must be carefully considered for population-level analyses: Simply optimizing for the best constraints on individual-planet atmospheric parameters will not necessarily lead to accurately and/or precisely constrained population parameters. Overall, there was no 1-to-1 mapping between the sample that had best constrained atmospheric parameters across all 12 planets and the sample that ultimately had the best constrained population parameters.

It is important to reiterate that 
five QS sample targets overlap with the ESBE sample, 
seven 3Bin sample targets overlap with the ESBE sample, and 
five QS sample targets overlap with the 3Bin sample. Because only one draw was done per planet per trial, differences between the success of the hypothetical survey (e.g. ESBE vs 3Bin) are driven by the planets that differ between the surveys. In the case of ``ESBE vs. 3Bin'' this would be just the difference between 
five out of 12 planets. These overlaps in the samples 
strongly motivate increasing the parent sample from which planets for this type of survey could be selected (i.e., those with mass constraints).

\section{Discussion \& Conclusions} \label{sec:summary}

In this work, we set out to investigate the role of sample selection in recovering population-level trends in exoplanet atmospheres. To do so, we create three hypothetical 12-planet surveys using three different methods: 1) quantitative selection (QS), where the planets are chosen via a purely quantitative merit function, 2) three-bin selection (3Bin), where the planets are chosen by picking the best four targets per bin in three radius bins, and 3) evenly-spaced-by-eye selection (ESBE), where the planets are chosen based on the subjective qualification that they adequately span the relevant parameter space (and meet a $t_{\rm{exp}}$ cutoff).

As our test case, we inject a population relation between the probability of a planet's atmosphere having a C/O ratio below 1 and the planet's radius and insolation flux. 
We emphasize that the injected relation is not meant to represent a physically probable model 
but instead serves as a vehicle for drawing broader conclusions about how sample selection will affect population-level studies.

For individual atmospheric targets, we find that the 3Bin- and the QS-derived samples result in only slightly more accurately and precisely retrieved atmospheric parameters across 10 trials, as compared to the ESBE sample. Thus, if population parameters were not of interest, any one of these sample results would be an acceptable strategy for choosing JWST targets. Additionally, this finding is verification that the 
results of this analysis are not biased by considering a sample selection method that does not result in suitable targets for atmospheric studies. Each sample has equal opportunity to succeed with regards to constraining the individual planets' atmospheric properties. 
Nevertheless, we find that there is not a one-to-one mapping between the samples with the better atmospheric targets and the samples with the best-constrained population parameters. Therefore, simply optimizing for the best targets for atmospheric characterization will not necessarily lead to robustly constrained population parameters. 

With regards to the population parameters, ESBE- and QS-derived samples resulted in the most accurate and precise parameters across all 10 trials, based on our population-level metric. When a uniform prior was chosen for the population level parameters, the ESBE-derived sample resulted in more trials with the overall highest accuracy population parameters, whereas the QS-derived sample resulted in the overall highest precision population parameters. When the choice of prior was changed, the ESBE-derived sample resulted in more trials with overall highest precision population parameters and the QS-derived sample resulted in overal highest accuracy. Therefore, without considering any other external factors, these two sample selection methods would be equally suitable. However, there are  two more factors that are also important to consider when choosing a sampling method.

First, the total observing charge time required to conduct all three surveys must be taken into account. The method for constructing the ESBE- and 3Bin-sample relies on well-sampling a parameter space either ``by-eye'' or ``within bins''. The QS-derived sample does not have as stringent of a requirement, instead relying on a ranking function (albeit designed to span a parameter space) to dictate target selection. Therefore, QS had more flexibility to choose targets that required less total observing time, even if they were relatively close in parameter space to another optimal target.  The QS-derived sample (94.2 hours) was approximately equal in efficiency to the ESBE-derived sample (89.7 hours) and both were more efficient than the 3Bin-derived sample (106.9 hours). 

Second, we must consider the initial motivation for reducing bias in constructing samples. It is important to account for how certain targets are chosen over others in order to accurately infer population parameters. Population studies are not meaningful if the inferred result changes depending on what fraction/subset of the underlying population is included in the sample. Without a reproducible (quantified) selection process, it is impossible to know to what degree the sample is affecting the results, let alone correct for that selection in quantitative population analyses. Although the ESBE sample appears to span the relevant parameter space and does perform well in some of our metrics, creating a sample that is truly evenly spaced is challenging in most cases, is subject to human bias, and by definition is not reproducible. For instance, the filters that we applied to create the parent sample for ESBE resulted in a total of 30 possible targets,
leaving many possible alternate combinations of targets that would still be ``evenly spaced by eye''. It is difficult to know if these various combinations would reproduce the results of this study.

Based on this analysis, our final conclusions are:

\begin{enumerate}
    
    
    \item All three sampling methods explored here (quantitative selection, evenly spaced by eye, and three-bin selection) offer the same opportunity for obtaining robust constraints on individual planet's atmospheric parameters. The QS- and the 3Bin-derived samples produced slightly better targets for independent atmospheric analyses.
    
    \item There is not a one-to-one mapping between the samples with the better atmospheric targets and the samples with the best-constrained population parameters. Therefore, simply optimizing for the best targets for atmospheric characterization will not lead to the highest chance of successfully constraining population parameters. The method of sample selection must be considered.

    \item The quantitative-selection (QS) and the evenly-spaced-by-eye (ESBE) methods for sample selection produced the best constraints on population parameters.  However, we caution that the ESBE method is more susceptible to human bias as compared to the quantitatively-derived method.

    \item A strictly ``binned'' approach (3Bin) is not recommended for robust population analyses, even though it may produce suitable targets for individual atmospheric studies. For example, across all trials the  1$\sigma$ credible interval on the radius population parameter was significantly larger for the 3Bin sample when compared to the QS sample. For the ESBE sample, the same was true for all but one trial. 
    
    \item A quantitatively-derived (QS) sample offers flexibility in target selection when compared to a ``binned'' or ``by-eye'' approach. For example, a binned method has strict requirements as to where targets must fall in parameter space. A quantitative method that is ranking-based does not. This is important when the sample size to draw from is small, as is the case for super-Earths/sub-Neptunes with well-constrained masses.

    \item There may be large variability in population-level results with a sample that is small enough to fit in a single JWST cycle ($\sim$12 planets), suggesting that the most successful population-level analyses will be multi-cycle.
  

\end{enumerate}

\acknowledgments
This research has made use of the Exoplanet Follow-up Observation Program website \citep{exofop} and NASA Exoplanet Archive, both of which are made available by the NASA Exoplanet Science Institute at IPAC, which is operated by the California Institute of Technology under contract with the National Aeronautics and Space Administration (NASA). Support for this work was provided by NASA through grant 80NSSC19K0290 to JT, through NASA’S Interdisciplinary Consortia for Astrobiology Research (NNH19ZDA001N-ICAR) under award number 19-ICAR19\_2-0041 to NMB and NEB, and from STFC grant ST/W507337/1 and University of Bristol School of Physics PhD Scholarship Fund to LA. Co-Author contributions are as follows: NEB helped conceptualize the motivation and analysis methodology, conducted the atmosphere simulations, retrievals, parameterized the results, contributed much of text, created Figures 2-7, and led the incorporation of coauthor comments. AW helped conceptualize the motivation and analysis methodology, developed and executed the statistical model and decided how to parameterize the results, and contributed text to the paper. JT helped conceptualize the motivation and analysis methodology, developed the sample selection strategies, and contributed text and Figure 1 to the paper and helped incorporate coauthor comments. MA, LA, MLM, NMB, and HRW provided detailed comments that greatly improved the quality and clarity of the paper. 
We thank the anonymous referee for their comments that helped improve the quality and clarity of this paper.

%



\software{PICASO \citep{picaso2020},  
          dynesty \citep{dynesty}, 
          pyStan \citep{Stan2019},
          PandExo \citep{batalha2017pandexo}
          }






\bibliographystyle{aasjournal}



\end{document}